# Shear Melting and High Temperature Embrittlement: Theory and Application to Machining Titanium

Graeme J Ackland[1] Con Healy[1] Sascha Koch[1] Florian Brunke[2] and Carsten Siemers[2]
[1] School of Physics, University of Edinburgh, Edinburgh EH9 3FD Scotland, UK.
[2] Technische Universitaet Braunschweig. IfW, Langer Kamp 8, 38106 Braunschweig, Germany

## ABSTRACT

We show that alloying with rare earth metals (REMs) can dramatically improve the machineability of a range of titanium alloys, even though the REM is not incorporated in the alloy matrix. The mechanism for this is that under cutting, shear bands are formed within which the nano-precipitates of REM are shear mixed. This lowers the melting point such that the mechanism of deformation changes from dislocation mechanism to localised amorphisation and shear softening. The material then fractures along the thin, amorphous shear-band. Outside the shear band, the REM remains as precipitates. The new alloys have similar mechanical properties and biocompatibility to conventional materials.

## INTRODUCTION

The relatively high cost of titanium alloy components limits their use in mass product applications. In the automotive industry, for example, steel is used even if $CO_2$-emissions could be reduced by using lighter titanium components. The high price is a result of the intrinsic raw material costs of titanium, manufacturing costs to produce a semi-finished part and the machining costs. For complex shaped products, machining of titanium alloys can account for about 50% of the product costs (Tönshoff & Hollmann, 2005).

Machining of titanium alloys is difficult because of the physical, chemical and mechanical properties of titanium (Lütjering & Williams, 2007). In addition, machining operations like drilling and turning in many cases cannot be automated due to the formation of long chips which wrap around the tools (Siemers, Laukart, Zahra, Rösler, Spotz & Saksl, 2010). Therefore, the cutting process has to be frequently interrupted, as it is necessary to remove the chips from the process zone by an operator, otherwise the finished surfaces and/or the tools might be destroyed. In titanium machining, three different kinds of chips are known to form: continuous chips having a constant chip thickness over the chip's length, segmented chips showing a saw-tooth like structure and completely separated segments if extreme cutting conditions are applied (Hou & Kommanduri, 1997; Obikawa, Anzai, Egawa, Narutaki, Shintani & Takeoka, 2011).

In case of segmented chip formation in titanium alloys, the process starts with damming of the material in front of the tool tip. During further progress of the tool, the deformation in the primary shear zone (a small area leading from the tool tip to the uncut surface) localizes (Healy, Koch, Siemers, Mukherji & Ackland, 2015). Finally, a segment is formed by shear deformation along a shear plane forming a so-called adiabatic shear band with a width of a few micrometers. During the shear band formation in titanium alloys local temperatures can raise to more than 1000°C, the plastic deformation can easily exceed 800% and the material becomes amorphous (Healy et al., 2015).

Extended machining studies using more than twenty different titanium alloys applying orthogonal as well as standard cutting operations have been carried out. A wide range of cutting speeds, cutting depths and feed rates has been applied (Siemers et al., 2010; Siemers, Jencus, Baeker, Roesler & Feyerabend, 2007). We show that titanium alloys containing high amounts of α-phase (like CP-Titanium Grades 2 and 4, α- and near-α-, (α+β)- as well as aged metastable β-alloys) form segmented chips for almost all machining processes and for a wide range of cutting speeds and cutting depths. Solution treated metastable β- as well as near-β- und β-alloys on the other hand show a cutting parameter dependent change in the chip formation process from continuous to segmented chips (Siemers, Laukart, Roesler, Rokicki & Saksl, 2011).

**RESULTS: THEORY**

**<u>Molecular dynamics of shear localisation and melting</u>**

Due to shear banding, the effective shear rates within the band may be enhanced by several orders of magnitude relative to the nominal shear rate of the entire sample. Such rates can be explored at the atomistic level using molecular dynamics simulations. By contrast, the very small thickness of the band makes finite-element modelling unfeasible due to small grid size. We have simulated regions of nanocrystal within the shear band to determine the mechanism of strain at extremely high shear rates.

Calculations were carried out using the MOLDY molecular dynamics code (Ackland et al., 2011) with up to 1,024,000 atoms in a polycrystalline bcc structure, and are similar to those reported in previous work on nanopillars (Healy and Ackland, 2015), except that here we used periodic boundary conditions to describe a region within a shear band. The polycrystal was created by creating regions of several hundred atoms in a bcc arrangement, and placing them with random orientations within holes cut from a liquid state simulation. On cooling this starting configuration to below the melting point, the nuclei were able to grow and for grain boundaries of random orientation when they encountered one another. This process also guarantees that the atom density in the boundary is appropriate. The polycrystalline sample was the subjected to uniaxial shear rates up to $5 \times 10^8$ s$^{-1}$ at temperatures up to 1600K. We found that the yield stress was relatively constant up to 900K, but beyond that it dropped markedly. Figure 1 shows visualisation of the simulation which reveals that at low temperatures the mechanism is a combination of dislocation motion and grain boundary motions, whereas above 900K a dynamics annealing of the microstructure occurred, followed by a localization of the shear in a highly disordered, melt-like region. A most striking result at the highest temperature was the spontaneous localization of the shear within the simulation region – a shear band just 10nm across. Analysis of common neighbours or displacement confirms that this high-shear region is non-crystalline.

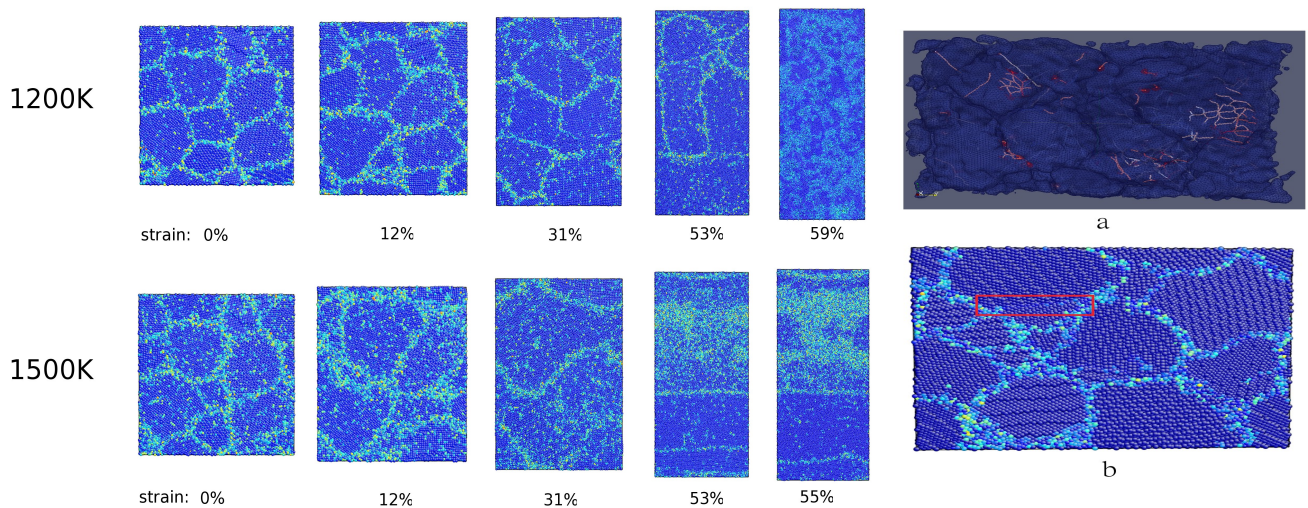

Figure 1 – MD simulations of shear at (left) high-T simulation imaged with centrosymmetry analysis in AtomEye – darker (blue) atoms are crystalline (bcc) and lighter (white) atoms are not. Initially shear procedes by grain boundary sliding, but above 50% a three dimensional region is amorphised, and in the 1500K case this shear band becomes localised. (right) at 300K, the same initial sample tranforms by (a) dislocations, identified by DXA algorithm (Stukowski and Albe, 2010) or (b) twinning, as visible in the box.

### Theoretical Interpretation and extrapolation of results

The picture which emerges from the simulation is that close to the melting point, there will be a transition from dislocation-driven shear to localised melting with very low yield stress. Such a region will enable the sample to break under high shear – for example during machining. We have found a mechanism by which titanium chips can break.

Key to this is to obtain temperatures close to the melting point. Although machining is carried out at room temperaure, the shearing process releases significant amounts of energy within the shear band, sufficient to raise the local temperature within the shear band by several hundred degrees. This leads to a feedback effect where the material softens, allowing the shear band where the energy is being produced to become thinner. Heating in a smaller area makes the sample still hotter: calculations balancing heat flows and energy production give an estimated temperature around 1300K, well below the melting point of standard titanium alloys. What is required is to lower the melting point of the material.

## RESULTS: EXPERIMENT
### Alloy Production Microstructure and Phase Composition

We produced alloys in a laboratory-size plasma-beam cold-hearth melting facility (PBCHM) based on Ti 6Al 4V, which is already used in automotive applications e.g. as fastener material. Rare earth additions were made to the standard alloy: Ti 6Al 4V 0.9La/1.5La, Ti 6Al 4V 0.9Ce, Ti 6Al 4V 0.9Er. The effect of La on a range of alloy compositions, potentially appropriate for various applications were also tested in our study: Vanadium has partly been replaced by different ferromolybdenum pre-alloys or niobium to improve the ductility or biocompatibility. In addition, small amounts of copper (known to enhance the deformability of titanium (Peters & Leyens, 2002)) and silicon (improving the castability as well as the oxidation and creep behavior (Lütjering & Williams, 2007)) were added to particular alloys. These additions do not affect the phase stability of primarily beta-phase alloys (Tegner at al). Finally, the alloys with rare-earth metals were : Ti 6Al 7Nb 0.9La, Ti 6Al 2Fe 1Mo 0.9La, Ti 6Al 1Fe 2Mo 0.9La, Ti 6Al 2Fe 1Mo 0.9La 0.5Cu and Ti 6Al 2V 3Nb 0.9La 0.7Fe 0.3Si.

The solubility of REMs in titanium is negligible at room temperature. The microstructure of all REM-containing alloys after melting, casting and a solution treatment consisted of a titanium α-matrix and discrete, equiaxed, metallic REM particles (identified by synchrotron diffraction) with a diameter between 2 μm and 20 μm in all 0.9% REM-containing alloys. Higher REM contents lead to a decoration of the grain boundaries. The initial grain size was similar in all cases and lay between 50 μm and 150 μm. Thermo-mechanical treatments applied to the alloys neither changed the particle size nor the location of the particles in the microstructure. Therefore, the particle dissolution kinetics must be very slow as the microstructure did not change even after eight-hour heat treatments 200°C above β-transus temperature.

In most of the alloys, the particles were located mainly on the grain boundaries which can be explained as follows: The plasma during arc melting superheats the melt by about 30°C to 50°C only (Peters & Leyens, 2002). The melting point of titanium (1668°C) and the investigated alloys (between 1630°C and 1705°C) is higher than the melting temperature of REMs, e.g. $T_{M,Ce}$ = 798°C, $T_{M,La}$ = 918°C and $T_{M,Er}$ = 1529°C, used. During cooling, the crystallization of the titanium matrix (containing the alloying elements) starts at the edges of the mould, grains form with only a limited amount of REM dissolved in titanium. The remaining liquid phase must therefore be enriched in REMs. Finally, once the matrix is fully crystallized, the remaining REM trapped on the grain boundaries and crystallizes during further cooling.

## Machinability and Mechanical Property Tests

During turning of Ti 6Al 4V, long chips formed as expected. Erbium-containing alloys also formed long chips, but when pure metallic cerium or lanthanum were present in the alloys, short chips were formed. An industrial ingot based on Ti 6Al 4V has been used in the production of mouthpieces for brass instruments as shown in Figure 2. The drill requires the use of step-drillers and conical reamers which is extremely difficult in conventional alloys due to long chip formation. These problems are overcome by of free-machining titanium alloys.

At room temperature, the tensile strength of the lanthanum containing Ti 6Al 4 0.9La alloy was about 8% higher than the standard alloy, whereas the ductility was diminished by about 30%. On the one hand, the grain refinement caused by the lanthanum particles led to strengthened alloys. On the other hand, the particles weakened the grain boundaries and intergranular cracks could propagate more easily. Even so, the yield tensile strength, the ultimate tensile strength and especially the minimal elongation at rupture (Ti 6Al 4V 0.9La: 10.5%) of the industrially produced material fulfilled the requirements of ASTM B348 for Ti 6Al 4V alloy. The Ti 6Al 4V 0.9La industrially processed material showed a fatigue limit of about 550 MPa (at R ≈ 0.01). The lanthanum particles had a negative effect on the fatigue properties as they acted as crack initiating points. The fatigue limit of the standard alloy Ti 6Al 4V was 600 MPa (at R ≈ 0.01) which was about 10% higher compared to the lanthanum containing alloys. Biocompatibility, corrosion resistance and the physical properties were unaffected (Feyerabend,

Siemers, Willumeit, Rösler, 2009). Even if Ti 6Al 2V 3Nb 0.9La 0.7Fe 0.3Si showed better mechanical properties, i.e. higher ductility (minimal elongation at rupture: 13%), compared to Ti 6Al 4V 0.9La, all the lanthanum modified alloys should preferably be applied in non-safety critical applications, e.g. in the automotive industry.

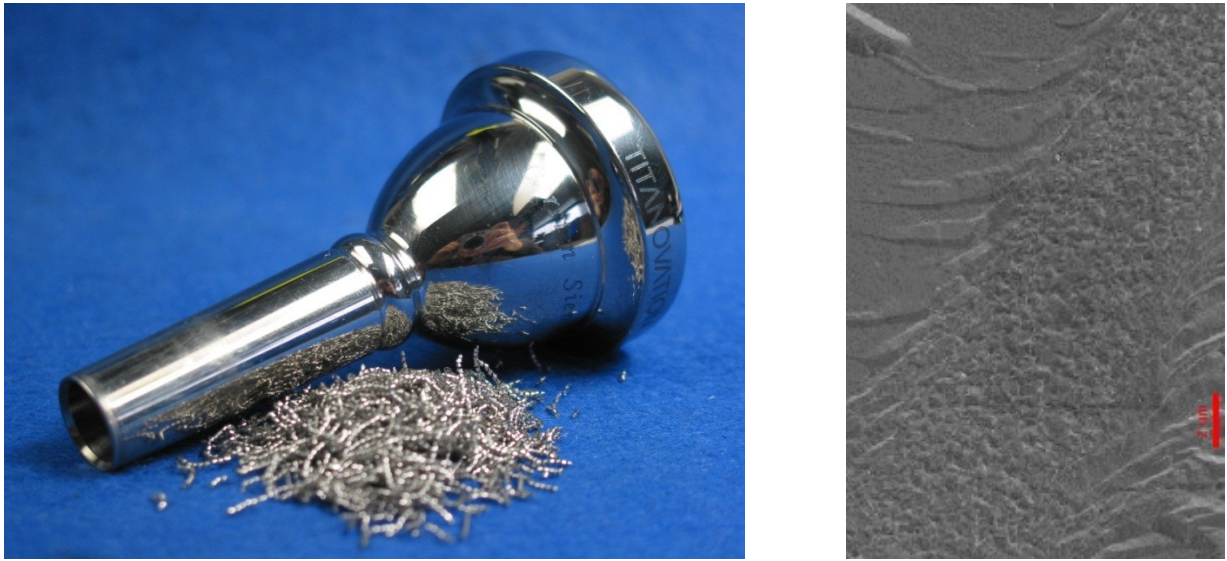

Figure 2 – (left) Mouthpiece for a bass trombone made of Ti 6Al 4V 0.9La, together with the short-breaking chips. (right) Microscopy image of a shear band in a conventional alloy after very high speed cutting, forming segmented but not short-breaking chips, the nanocrystalline material within the band is evidence of the amorphization, but for non-REM alloys complete fracture is not observed. Due to fracture, it was not possible to image the shear band in the REM-containing alloys.

**CONCLUSIONS**

Our simulations show that at high shear rates the grain structure is destroyed and elements can be forcibly shear-mixed. In all cases the stable crystal phase of this well-mixed alloy is bcc, unless melting occurs. During machining, the temperature in the shear bands between segmented chips reaches or exceeds 1000°C (Siemers, Bäker, Mukherji & Rösler, 2003), and the grain structure is destroyed by the high shear, forcibly mixing the REM into the matrix forming a composition above the eutectic melting point. Post-hoc analysis of unbroken shear bands shows a rapidly recrystallised nanocrystalline microstructure, supporting the eutectic - melting hypothesis.

The chips of the lanthanum containing alloys were separated in the primary shear zones (Rösler, Bäker, Siemers, 2008), implying that the loss of strength in the shear bands between segments causes chips to fall apart, either directly once the shear-softened shear band forms, or due to vibrations during further progress of the tool. The melting point of erbium and its eutectic with Ti is higher than the temperatures reached during the shear band formation, so the shear softening and associated fracture does not occur.

This technique for creating free-machining alloys was successfully tested on a range of representative alloys with compositions tailored to varied applications. The machineability of an industrially produced ingot of the standard Ti64+La alloy was demonstrated using a step-drill and conical reamer.

## ACKNOWLEDGMENTS

This work was produced under the EU-FP7-ITN project MAMINA.

## REFERENCES

Ackland,G. J., D'Mellow, K., Daraszewicz,S. L., Hepburn, D. J., Uhrin, M. and Stratford, K. (2011) The MOLDY short-range molecular dynamics package *Computer Physics Communications* 182, 2587.

Feyerabend, F., Siemers, C., Willumeit, R & Rösler, J. (2009). Cytocompatibility of a free machining titanium alloy containing lanthanum, *Journal of Biomedical Materials Research* A, 90A, 3, 931–939.

Healy, C., & Ackland, G.J. (2015). Molecular dynamics simulations of compression–tension asymmetry in plasticity of Fe nanopillars *Acta Materialia* 70, 105-112

Healy, C., Koch, S, Siemers, C., Mukherji, D. & Ackland, G.J. (2015). Shear melting and high temperature embrittlement: Theory and application to machining titanium, *Physical Review Letters 114 (16)*, 165501.

Hou, Z.B. & Komanduri, R. (1997). Modelling of thermomechanical shear instability in machining. *International Journal of Mechanical Sciences,* 39 (11), 1273-1314.

Li, J. (2003) Atomeye *Modelling Simul. Mater. Sci. Eng*. 11 173

Lütjering, G. & Williams, J.C. (2007). *Titanium*. Berlin, Germany: Springer.

Obikawa, T, Anzai, M., Egawa, T., Narutaki, N., Shintani, K. & Takeoka, E. (2011). High Speed Machining: A Review from a Viewpoint of Chip Formation. *Advanced Materials Research*, 188, 578-583.

Peters, M. & Leyens, C., Eds. (2002). *Titanium and titanium alloys*. Weinheim, Germany: Wiley-VCH.

Siemers, C., Laukart, J., Zahra, B., Rösler, J., Spotz, Z., & Saksl, K. (2010). Development of advanced and free-machining titanium alloys. In: D. Gallienne, M. Bilodeau (Eds.) *Fortynineth Conference of Metallurgists, Section Light Metals 2010 - Advances in Materials and Processes* (pp. 311-322), Vancouver, Canada: The Canadian Institute of Mining, Metallurgy and Petroleum.

Siemers, C., Jencus, P., Baeker, M., Roesler, J. & Feyerabend, F. (2007). A new free machining titanium alloy containing lanthanum. In: M. Niinomi, S. Akiyama, M. Hagiwara, M. Ikeda & K. Maruyama (Eds.), *Eleventh World Conference on Titanium* (pp. 709-712). Kyoto, Japan: The Japan Institute of Metals.

Siemers, C., Bäker, M., Mukherji, D. & Rösler, J. (2003). Microstructure evolution in shear bands during the chip formation of Ti 6Al 4V. In: G. Lütjering & J. Albrecht (Eds.). *Tenth World Conference on Titanium* (pp. 839-846). Hamburg, Germany: Wiley VCH.

Stukowski, A. and Albe, K (2010) Extracting dislocations and non-dislocation crystal defects from atomistic simulation data *Modelling and Simulation in Materials Science and Engineering*, 18, 085001

Tegner, B. E., Zhu, L., & Ackland, G. J. (2012). Relative strength of phase stabilizers in titanium alloys. *Physical Review B*, 85, 214106.